\documentclass[10pt,a4paper]{article}

\usepackage{siunitx}
\usepackage[T1]{fontenc}
\usepackage{multicol}
\usepackage{multirow}
\usepackage{rotating}
\usepackage{lscape}
\usepackage{array}
\usepackage{mwe}
\usepackage{graphicx}
\usepackage{ragged2e}
\usepackage{authblk}
\usepackage{rotating}
\newcommand{\cmmnt}[1]{\ignorespaces}

\title{Partition potential for hydrogen-bonding in formic acid dimers \\ \small (accepted for publication in the International Journal of Quantum Chemistry)}
\date{}

\author[1]{Sara G\'omez}
\author[2]{Yan Oueis}
\author[3]{Albeiro Restrepo}
\author[2,4]{Adam Wasserman}

\affil[1]{Departamento de Educaci\'on y Ciencias B\'asicas, Instituto Tecnol\'ogico Metropolitano, Calle 73 No. 76A--354, Medell\'in, Colombia}
\affil[2]{Department of Chemistry, Purdue University,  560 Oval Drive, West Lafayette, Indiana 47907, USA}
\affil[3]{Instituto de Qu\'imica, Universidad de Antioquia UdeA, Calle 70 No. 52-21, Medell\'in, Colombia.}
\affil[4]{Department of Physics and Astronomy, Purdue University, 525 Northwestern Avenue, West Lafayette, Indiana 47907, USA}

\begin{document}

\maketitle

\begin{abstract}
The ground-state energy and density of four low-energy conformations of the formic acid dimer were calculated via Partition Density Functional Theory (PDFT). The differences between isolated and PDFT monomer densities display similar deformation patterns for primary and secondary hydrogen bonds among all four dimers. In contrast, the partition potential shows no transferable features in the bonding regions. These observations highlight the global character of the partition potential and the cooperative effect that occurs when a dimer is bound via more than one hydrogen bond. We also provide numerical confirmation of the intuitive (but unproven) observation that fragment deformation energies are larger for systems with larger binding energies.
\end{abstract}

\pagebreak

\section{INTRODUCTION}
Despite the tremendous advances in modern computational methods and availability of efficient and powerful hardware, chemical applications of quantum mechanics are still heavily limited by the system size. The need to overcome this limitation becomes more pressing with the increasing interest in chemical properties of large systems, stimulated by research in chemical biology and materials science. One common approach to this problem involves breaking down the system into fragments for which properties can be obtained at a lower computational cost. The total properties of the system are then obtained from the properties of the fragments corrected for the inter-fragment interactions. \cite{L68,KSANU99,KIANU99} Among modern embedding methods that can be used for this purpose \cite{C91,WW93,cr2,HPC11,HC11,SC16}, those that use the electronic density as the main variable \cite{JN14,NW14,WSZ15} have the advantage of simplicity and can be directly connected to Kohn-Sham DFT \cite{KS65}; and among density-based embedding methods, Partition Density Functional Theory (PDFT) \cite{wasserman_2010, NW14} has the advantage of producing localized fragment densities that facilitate the connection to traditional chemical concepts. \cite{GPL03,jctc2018}\\

PDFT is conceptually analogous to Kohn-Sham DFT. In PDFT, a system of interacting fragments is uniquely mapped onto a fictitious system of non-interacting fragments in a global (\textit{i.e.} same for all fragments) external potential. This partition potential, $v_p(\mathbf{r})$, is unique for a given set of fragments. Although there are infinitely many ways to partition a system, chemically relevant fragments are usually the most natural choice of partitioning as they enable meaningful chemical properties to be calculated. \cite{pdt} In the case of the formic acid dimer\cite{pccp2016,jpca2017,jctc2018_2} analyzed here, we choose two monomers as fragments and label them \textit{left} ($L$) and \textit{right} ($R$). \\

As opposed to most other density embedding methods that minimize the total energy \cite{JN14,KSGP15}, in PDFT, we search for a minimum of the sum of fragment energies:
\begin{equation}
 E_{f}[n_{L}(\mathbf{r}),n_{R}(\mathbf{r})]= E_{L}[n_{L}(\mathbf{r})]+E_{R}[n_{R}(\mathbf{r})]
 \label{eq:fragE}
\end{equation}
subject to the constraint that fragment densities sum to the total density of the system at each point in space, $\mathbf{r}$:
\begin{equation}
n_{tot}(\mathbf{r})=n_{L}(\mathbf{r})+n_{R}(\mathbf{r})
 \label{eq:totn}
\end{equation}
This constrained optimization can be replaced by the unconstrained optimization of the following functional:
\begin{equation}
G[n_{L}(\mathbf{r}),n_{R}(\mathbf{r}),v_p(\mathbf{r})]=E_{f}[n_{L}(\mathbf{r}),n_{R}(\mathbf{r})]+\int d\mathbf{r}v_p(\mathbf{r})(n_{tot}(\mathbf{r})-n_{L}(\mathbf{r})-n_{R}(\mathbf{r}))
 \label{eq:G}
\end{equation}
In this formalism, the partition potential $v_p(\mathbf{r})$ appears as the Lagrange multiplier that controls the density constraint. More generally, PDFT can be formulated for varying non-integer fragment occupations. \cite{pdt, cr1, cr2} In this work, however, we fix occupations to the ones of isolated fragments. This simplification increases the efficiency of the method as optimization with respect to occupation numbers is not needed. Previous work on simpler systems suggests that occupations usually lock to integers when fragments have similar electronic structures. Since we work with neutral dimers with small dipole moments, we choose neutral fragments and focus attention on monomer density deformations: 
\begin{equation}
\Delta n_i(\mathbf{r})=n_i(\mathbf{r})-n_i^0(\mathbf{r})
\label{e:dn}
\end{equation}
where $i$ stands for either $L$ or $R$.

Among the many types of intermolecular interactions, hydrogen-bonding is of particular interest because hydrogen bonds (HBs) are known to be responsible for stabilization of various chemical systems, with direct implications in a wide range of scenarios, from the life-supporting properties of liquid water \cite{w4, w5, w6, w7} to the tertiary structures of biomolecules in charge of storing and replicating genetic information. \cite{bio_book} Although the very nature of hydrogen bonding is not without controversy \cite{wein1,aw1,grabowski, buckingham}, several types of HBs are recognized in the literature. \cite{gilli} Of particular interest to this work are the conventional \textit{primary hydrogen bonds}, where a hydrogen atom sits between two electronegative atoms, and the non-conventional, \textit{secondary hydrogen bonds}, where a proton is donated from a non-polar C--H bond.\\

Although PDFT is particularly well suited to study molecular clusters, these systems are challenging because intermolecular interaction energies in clusters are significantly smaller than energies associated with formal bonds. Individual molecules in clusters retain their chemical identities to a large degree and require carefully constructed partition potentials to account for the comparatively weak interactions. PDFT was recently successfully applied to water dimers, \cite{w2-pdft} where it was shown that the partition potential and PDFT densities can be used to describe the mechanism of hydrogen-bond formation. \\

The question we address in this paper is whether the partition potential has transferable features corresponding to particular types of HBs. Finding transferability would imply that the partition potential around a hydrogen bond in one molecule could be used as a starting point to calculate approximate interaction energies in other molecules with similar HBs, an appealing prospect for computational chemistry. The formic acid dimers, (FA)$_2$, are ideal systems to investigate this question because their four lowest-energy conformations have two types of primary HBs (C=O$\cdots$H--O and H--O$\cdots$H--O ) and two types of secondary HBs (C=O$\cdots$H--C, H--O$\cdots$H--C). Is $v_p(\mathbf{r})$ in the vicinity of a primary HB in one of these four dimers a good approximation to $v_p(\mathbf{r})$ for a primary HB in a different dimer?  What about the same question for secondary HBs? Previous work on one-dimensional model systems \cite{ZW10} taught us that the transferability of PDFT \textit{densities} was about an order of magnitude higher than that of real-space partitioning schemes, so it is reasonable to expect transferable features in the underlying partition potentials. However, we find that the answer is \textit{no} in both cases (primary and secondary), contrary to naive intuition. Conversely, monomer density deformations \textit{do} have specific features that can be used to distinguish between different types of HBs. \\

\section{LOW-ENERGY STRUCTURES OF FORMIC-ACID DIMERS}

\begin{figure}[h]
\centering
\includegraphics[scale=0.8]{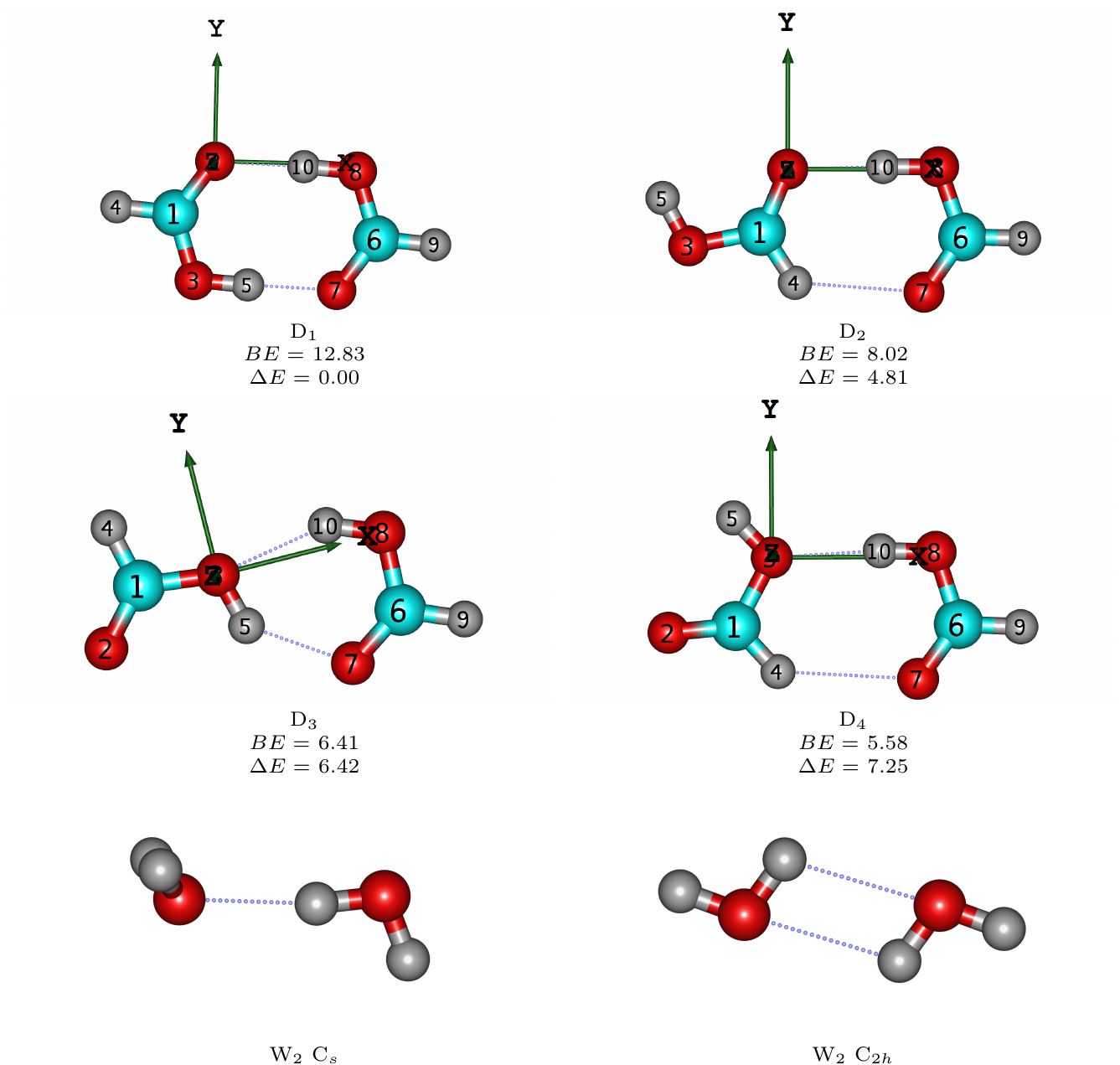}
\caption{Lowest energy dimers of formic acid from the MP2/6--311++G($d,p$) Potential Energy Surface. \cite{fad} Both monomers are in the $anti$ conformation. BEs are the CCSD(T)/6--311++G($d,p$) binding energies calculated as the difference between the given dimer and the isolated fragments. Relative energies with respect to the global minimum are shown. All energies in kcal/mol and corrected for zero-point vibrational energies. The right monomer ($R$), which simultaneously acts as a donor and acceptor of hydrogen bonds, is shown in the same perspective in all cases. The origin of coordinates is placed at the oxygen atom in the left ($L$) monomer. The two conformations of the water dimer discussed in the text are presented.}      
\label{f:structures}
\end{figure}

Despite being the smallest carboxylic acid, the conformational space for the formic acid dimers is notoriously rich, with a considerable number of structures already experimentally detected. \cite{lundell} Farf\'an \textit{et al.} \textit{et al.} calculated a set of 21 well-defined minima in MP2/6-311++G$(d,p)$ potential energy surface (PES). \cite{fad} For our work, we selected four lowest energy motifs from this set (shown in Figure\ref{f:structures}). We reoptimized their geometries and confirmed the found stationary points are true minima by frequency calculations using the B3LYP XC functional and Dunning's aug-cc-pVTZ basis set. PDFT calculations were performed over the resulting geometries. B3LYP and PW91 were used as XC functionals in the construction of the effective potentials. The partition potential was expanded using aug-cc-pVTZ basis set. PW91 has been shown to be useful for the evaluation of intermolecular interactions in hydrogen-bonded systems. In particular, the dimers of water and formic acid for which PW91 computed interaction energies showed only slight changes with respect to CCSD(T). \cite{tsuzuki} It has also been concluded that large basis sets for $v_p\left(\mathbf{r}\right)$ lead to accurate total densities. \cite{w2-pdft} All calculations were carried out using the NWChem package. \cite{nwchem} \\

\section{RESULTS AND DISCUSSION}

\subsection{Energy Analysis}
PDFT describes fragment interactions by means of two energy contributions: the preparation energy, $E_{prep}$, and the interaction energy among the "prepared" fragments (partition energy, $E_p$). The preparation energy is defined as the energy required to distort the density of isolated fragments to the density of fragments within the dimer, satisfying the condition of eq. \ref{eq:totn} and minimizing $G$ of eq. \ref{eq:G}:
\begin{equation}
 E_{prep}=E_{f}[n_{L}(\mathbf{r}),n_{R}(\mathbf{r})]-(E_{L}[n_{L}^0(\mathbf{r})]+E_{R}[n_{R}^0(\mathbf{r})]),
\label{eq:eprep}
\end{equation}
where $n_i^0(\mathbf{r})$ is the density of an isolated $i$-fragment. We can also identify the preparation energy of a fragment $i$:
\begin{equation}
E_{prep}^{i} = E_{i}[n_{i}(\mathbf{r})]-E_{i}[n_{i}^0(\mathbf{r})].
\label{eq:eprepi}
\end{equation}
The partition energy describes the interaction of these distorted fragments and is defined as:
\begin{equation}
E_{p} = E_{Dimer}[n_{L}(\mathbf{r}),n_{R}(\mathbf{r})]-E_{f}[n_{L}(\mathbf{r}),n_{R}(\mathbf{r})].
\label{eq:epart}
\end{equation}
$v_{p}(\mathbf{r})$ is the functional derivative of $E_{p}$ with respect to the total density. In our calculations, however, $v_p(\mathbf{r})$ and $E_p$ are calculated separately:  $v_p(\mathbf{r})$ through a numerical inversion and $E_{p}$ by eq. \ref{eq:epart} from the resulting energies. The inversion algorithm is based on computing the fragment density response, $\chi_f(\mathbf{r,r^\prime})$ at each step and updating $v_p(\mathbf{r})$ through: \cite{w2-pdft,WY03}
\begin{equation}
\delta v_p(\mathbf{r}) = \int d\mathbf{r^\prime} \chi_f(\mathbf{r,r^\prime})^{-1} \left( \sum_i n_i(\mathbf{r^\prime}) - n_f(\mathbf{r^\prime}) \right),
\label{eq:WY}
\end{equation}
where $n_f(\mathbf{r})=n_L(\mathbf{r})+n_R(\mathbf{r})$. The binding energy, $BE$, is simply the sum of the two terms:
\begin{equation}
BE=E_{p}+E_{prep} = E_{Dimer}[n_{L}(\mathbf{r}),n_{R}(\mathbf{r})]-(E_{L}[n_{L}^0(\mathbf{r})]+E_{R}[n_{R}^0(\mathbf{r})])
\label{eq:be}
\end{equation}
Since all four dimers have their atoms lie on a plane, it is most convenient to visualize the densities and potentials at values of $\mathbf{r}$ belonging to this plane. \\

Table \ref{energies} lists relevant quantities extracted from the PDFT calculations using both B3LYP and PW91 functionals. The experimental BE for D$_1$ is -59.5 $\pm$ 0.5 kJ/mol (2.27 $\times$ 10$^{-2}$ a.u.) \cite{be_fad_exp}. The BEs computed with eq. \ref{eq:be}  yield 2.90 $\times$ 10$^{-2}$ a.u. (B3LYP)  and 3.14 $\times$ 10$^{-2}$ a.u. (PW91). We note that these values do not include zero-point vibrational energies and thermal corrections. \\

\begin{sidewaystable}
\centering 
\caption{Relevant energies (a.u.) from PDFT calculations (B3LYP, PW91, $v_p\left(\mathbf{r}\right)$ with the aug--cc--pVTZ basis set)  on the lowest energy formic acid dimers (Figure \ref{f:structures}). The energy for the isolated $anti$ formic acid monomer is $E_{L}^0=E_{R}^0=$-189.846 a.u. Analogous results for the water dimer in two different symmetries\cite{w2-pdft}  are also shown for comparison. All energies in kcal/mol.}
\label{energies} 

\begin{tabular}{|lllllll|} \hline \hline
System & $BE$ & $E_{prep}^{D_i}$ (Eq. \ref{eq:eprep})  & $E_{prep}^{L}$ & $E_{prep}^{R}$ & $E_{p}$ & Intermolecular Interactions \\
\hline \hline
D$_1$  \tiny{(B3LYP)} & -18.19 & 7.14 & 3.57 (50\%) & 3.57 (50\%) & -25.33 & \tiny C=O $\cdots$ H--O $\qquad$ C=O $\cdots$ H--O\\
D$_1$  \tiny{(PW91)}  & -19.70 & 6.99 & 3.49 (50\%) & 3.49 (50\%) & -26.69 & \tiny C=O $\cdots$ H--O $\qquad$ C=O $\cdots$ H--O\\

D$_2$  \tiny{(B3LYP)} & -9.61 & 4.67 & 2.56 (55\%) & 2.11 (45\%) & -14.28 & \tiny C=O $\cdots$ H--O $\qquad$ C=O $\cdots$ H--C   \\
D$_2$  \tiny{(PW91)}  & -10.55 & 4.76 & 2.76 (58\%) & 2.00 (42\%) & -15.31 & \tiny C=O $\cdots$ H--O $\qquad$ C=O $\cdots$ H--C   \\

D$_3$  \tiny{(B3LYP)} & -6.87 & 2.08 & 1.14 (55\%) & 0.94 (45\%) & -8.95 & \tiny C=O $\cdots$ H--O $\qquad$ H--O $\cdots$ H--O \\
D$_3$  \tiny{(PW91)}  & -7.84 & 2.08 & 1.11 (54\%) & 0.97 (46\%) & -9.92 & \tiny C=O $\cdots$ H--O $\qquad$ H--O $\cdots$ H--O \\

D$_4$  \tiny{(B3LYP)} & -5.01 & 1.71 & 1.06 (62\%) & 0.66 (38\%) & -6.72 & \tiny C=O $\cdots$ H--C $\qquad$ H--O $\cdots$ H--O\\
D$_4$  \tiny{(PW91)}  & -5.62 & 1.75 & 1.13 (65\%) & 0.62 (35\%) & -7.37 & \tiny C=O $\cdots$ H--C $\qquad$ H--O $\cdots$ H--O\\
W$_2$ C$_s$ \tiny{(B3LYP)} & -4.45 & 1.86 & 0.88 (47\%) & 0.98 (53\%) & -6.31 & \tiny H--O $\cdots$ H--O \\
W$_2$ C$_{2h}$ \tiny{(B3LYP)} & -3.09 & 0.42 & 0.21 (50\%) & 0.21 (50\%) & -3.50 &\tiny H--O $\cdots$ H--O $\qquad$ H--O $\cdots$ H--O\\

\hline
\end{tabular}
\end{sidewaystable}

An inventory of intermolecular interactions in the four formic acid dimers studied here is provided in the rightmost column of Table \ref{energies}. D$_1$ and D$_3$ are stabilized  by primary hydrogen bonds only while D$_2$ and D$_4$ include one secondary hydrogen bond each.  D$_1$ exhibits two equivalent primary hydrogen bonds where the hydroxyl group in one monomer donates a proton to the carbonyl group of the other (see Figure {\ref{f:structures}). D$_2$ and D$_4$ have two types of contacts: a secondary C=O $\cdots$ H--C hydrogen bond and a C=O $\cdots$ H--O (D$_2$) and H--O $\cdots$ H--O (D$_4$) primary HB. D$_3$ has two non--equivalent primary hydrogen bonds, where $R$ simultaneously acts as donor and acceptor in two different functional groups, the O--H bond in $L$ acts as donor and acceptor of both HBs, freeing the carbonyl group in $L$ of intermolecular interactions. We note that although secondary hydrogen bonds are typically considered weaker than primary hydrogen bonds \cite{gilli}, the overall stability of the dimers is not correlated with the primary or secondary nature of the HBs. For example, D$_2$ is lower in energy than D$_3$, even though D$_2$ has one primary and one secondary HBs and D$_3$ has two primary HBs. This lack of correlation extends to the number of hydrogen bonds, as seen for example in the two water dimers listed in Table \ref{energies}. \\

Table \ref{energies} also lists preparation energies for each dimer and its components according to eqs. \ref{eq:eprep} and \ref{eq:eprepi}. It is clear from eqs. \ref{eq:fragE} and \ref{eq:eprep} that $E_{prep}$ is always positive. We also expect larger values of $E_{prep}$ for fragments that are more distorted relative to their isolated states. $E_{prep}$ decreases in going from D$_1$ to D$_4$. Fig. \ref{f:dens_diff_plots}, which shows the densities on the plane of two monomers, makes it obvious that this decrease corresponds to the decrease in the total density deformation. \\

There are characteristic deformation patterns for the primary and secondary bonds, as shown in Fig. \ref{f:dens_diff_plots}. The O atom of the H-donating O--H group has a significant density increase along the approximate direction of the HB in a dumbbell-like shape. The O atom of the acceptor has a density decrease of similar shape and direction. The H atom of the O--H group also has some density deficiency around it. The secondary bond pattern is very similar (note that the O--H donor is now replaced with C--H), but the deformation is smaller in magnitude and is more disperse. These observations suggest that the stronger intermolecular bonds require larger deformation of the original wavefunctions of the fragments, a result that may appear obvious to many chemists, but can not be quantified without a rigorous definition of fragments within a molecule. This is also consistent with the orbital interaction picture where it is generally thought that the gain of electron density in the $\sigma_{\rm O-H}^{\star}$ region of the donor and the simultaneous loss of charge in the region associated to the O atom of the acceptor is responsible for the formation of  a hydrogen bond.  \cite{wein1, aw1, fad} Although fragment occupation numbers remain constant in the present PDFT implementation, fragment densities are indeed distorted; these distortions are linked to the charge transfer within fragments, provided by the orbital interactions. \\

The fragment preparation energies can be analyzed further. Since in D$_1$ both monomers are the same, their preparation energies are identical. In D$_2$ and D$_4$ the left monomer acts as a donor of a secondary hydrogen bond. In those cases, $E_{prep}^{L}$ is significantly larger than $E_{prep}^{R}$, even though the density deformation reaches higher values in the $R$-monomer. The $L$-monomer has a more delocalized density deformation. This imbalance can also be attributed to the fact that weaker secondary HBs require smaller preparation energies. \cmmnt{We attribute such imbalance to the reaccommodation of charge transferred to $\sigma^{\star}_{\rm C-H}$ orbitals in $L$, which seems to require larger energies than the ones needed to distort the geometries of $R$ to incorporate the transferred charge to the $\sigma^{\star}_{\rm O-H}$ orbitals.}In D$_3$, the energy needed to prepare $L$ is larger than the energy needed to prepare $R$ because of the double donor/acceptor function of the O--H group in $L$. \\

Partition energies, $E_p$, also shown in Table \ref{energies}, are always negative and their magnitudes are correlated with the corresponding $E_{prep}$'s. That $E_p$ is negative can be proven from the variational principle, but the observed correlation with $E_{prep}$ (\textit{i.e.} that $E_{prep}$ decreases as $E_p$ decreases) cannot. As predicted by the analysis of Fig. \ref{f:dens_diff_plots}, larger preparation energies lead to larger partition energies, which is seen for all dimers in Table \ref{energies}. This trend is followed not only by FA dimers but by all other systems we have studied so far. Whereas this observation seems obvious, a hard proof is missing. \\

\begin{figure}
\centering
\includegraphics[scale=0.6]{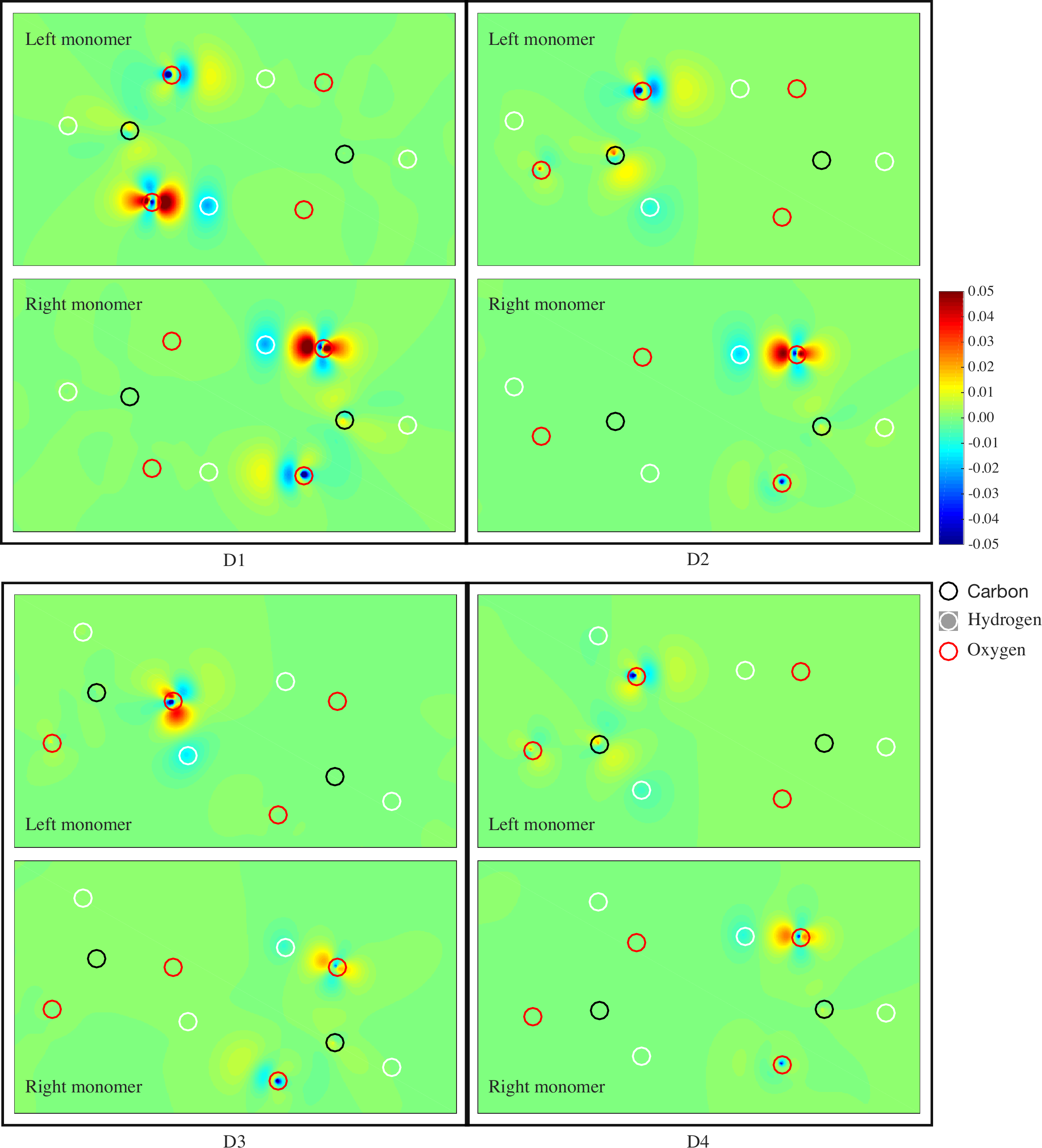} 
\caption{Density distortions, eq. \ref{e:dn}, On the molecular plane for various dimer configurations (in a.u.). For clarity, atom positions are indicated by hollow circles centered at atomic positions and bond lines are omitted. Upper panels correspond to the left monomer; lower panels correspond to the right monomers.}
\label{f:dens_diff_plots}
\end{figure}

It was noticed in previous work \cite{w2-pdft} that the character of chemical interactions (formal bonds, long range, van der Waals, etc.) appears to be related to the magnitude of the preparation energy. Thus, as expected, preparation energies in Table \ref{energies} suggest a direct correlation with binding energies. Table \ref{t:prep_energies} lists recalculated total $E_{prep}$ for a set of diatomic and polyatomic systems already available in the literature, as well as our results for the formic acid dimers. Nicely, the preparation energies for (FA)$_2$ fall in the same range as that of the water dimer. They lead to interaction energies that are stronger than van der Waals contacts but weaker than ionic and covalent bonds. In hydrogen bonding, the distance separating the two moieties dictates the strength of the interaction. This tendency can be seen in Table \ref{t:prep_energies}.

\begin{table}[h]
\caption{Total preparation energies for different systems. All PDFT calculations using B3LYP/aug--cc--pVTZ with an expansion of $v_p(\mathbf{r})$ in the same basis set. $R_{\rm O-O}$ and $R_{\rm C-O}$ are the distances between oxygen atoms in primary hydrogen bonds and between carbon and oxygen atoms in secondary hydrogen bonds, respectively.} 
\label{t:prep_energies} 
\begin{center}
\begin{tabular}{|c|cc|} 
\hline 
\hline
   System  & Distance (\AA{}) & $E_{prep}$ (kcal/mol)$^a$ \\
\hline
LiH (neutral fragments)& 1.59 & 34.76\\
LiH (ionic fragments)  & 1.59 & 23.59 \\
H$_2$                  & 0.74 & 12.76\\
D$_1$& $R_{\rm O-O} =$ 2.67   & 7.14 \\
D$_2$& $R_{\rm O-O} =$ 2.73, $R_{\rm C-O} =$ 3.13 & 4.67 \\
D$_3$& $R_{\rm O-O} =$ 2.73, 2.89 & 2.08 \\
$C_s$ Water Dimer& $R_{\rm O-O} =$ 2.86 & 1.86 \\
D$_4$& $R_{\rm O-O} =$ 2.90, $R_{\rm C-O} =$ 3.35 & 1.71 \\
$C_{2h}$ Water Dimer& $R_{\rm O-O} = $ 2.76 & 0.42 \\
He$_2$                 & 1.60 & 0.53 \\

\hline
\end{tabular}\\
\end{center}
$^a$Our results for the diatomic molecules in this table differ slightly from those of the original work of Nafziger, Wu, and Wasserman \cite{nafziger_2011} because we recalculated all energies using the aug--cc--pVTZ basis set. \\
\end{table}

\subsection{Partition potentials}

Hydrogen bonding is a complex interaction with various degrees of contribution from electrostatic, inductive, and dispersive forces depending on the nature of the individual molecules. In this work, we use the PW91 functional, which fortuitously yields accurate interaction energies and molecular geometries in weakly-bonded clusters such as the benzene and methane dimers among others \cite{tsuzuki}; and we also use the very popular B3LYP hybrid functional.\\

Fig. \ref{f:rho_vp_plots} shows that all features of the partition potential are largely insensitive to the choice of XC functional (we only show results for D$_1$, but the same is also true for D$_2$ $\rightarrow$ D$_4$). Since $v_p(\mathbf{r})$ is obtained through the density-to-potential inversion of eq. \ref{eq:WY}, this is due to the \textit{densities} being insensitive to the choice of XC functional. The question of whether the approximate XC functionals can accurately capture the exact features of $v_p(\mathbf{r})$ remains open. \cite{OW18} For both functionals, we were able to achieve density convergence to the order of $10^{-8}$ a.u. in a reasonable number of iterations (on the order of $10^2$). \\

Fig. \ref{f:vp_plots} compares the B3LYP partition potentials for all four dimers. In contrast to the monomer density deformations, similar bonds are \textit{not} characterized by similar features in $v_p(\mathbf{r})$. This is further highlighted in Figure \ref{t:rho_vp_d2_d3_d4}, where the partition potentials are plotted along the following nearly-linear intramolecular bonds: C=O $\cdots$ O--H in D$_1$ and D$_2$; O--H $\cdots$ O--H in D$_3$; and O--H $\cdots$ O--H in D$_4$. Note that although the density deformations in the binding regions are qualitatively similar in all four dimers, $v_p(\mathbf{r})$ is qualitatively different for the global minimum (D$_1$), where it is highly negative. \\

The non-transferability of $v_p(\mathbf{r})$ or any of its features indicates its sensitivity to the density variations in regions that may be far from $\mathbf{r}$. In contrast, the density deformations are highly localized. Qualitatively, the 2D density deformations  of Figure \ref{f:dens_diff_plots} show that, due to the formation of the hydrogen bond, charge is accumulated in the region occupied by the antibonding orbital in the $R$ monomer and simultaneously withdrawn from the region occupied by the lone pair in the $L$ monomer. Quantitatively, accumulation of charge in the antibonding region of $R$ and depletion of charge in the lone pair region of $L$ are larger for stronger bonds.
\begin{figure}
\centering
\includegraphics[scale=0.2]{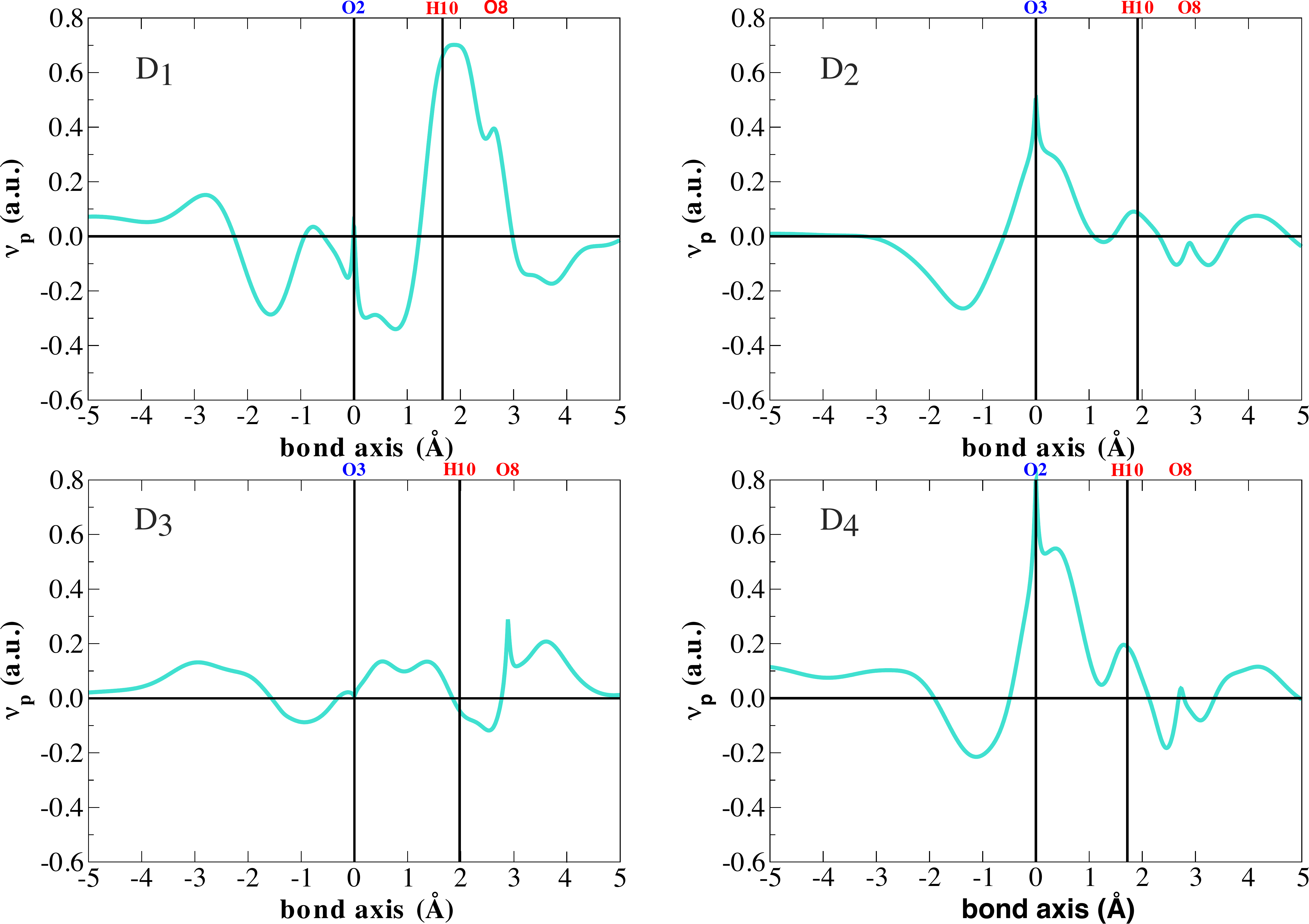} 
\caption{1D plots of $v_p(\mathbf{r})$ along the approximate bond axis. The B3LYP functional in conjunction with the aug-cc-pVTZ basis set was used for all calculations. Vertical lines in the 1D plots enclose the intermolecular bonding region.}
\label{t:rho_vp_d2_d3_d4} 
\end{figure}
\begin{figure}
\centering
\includegraphics[scale=0.60]{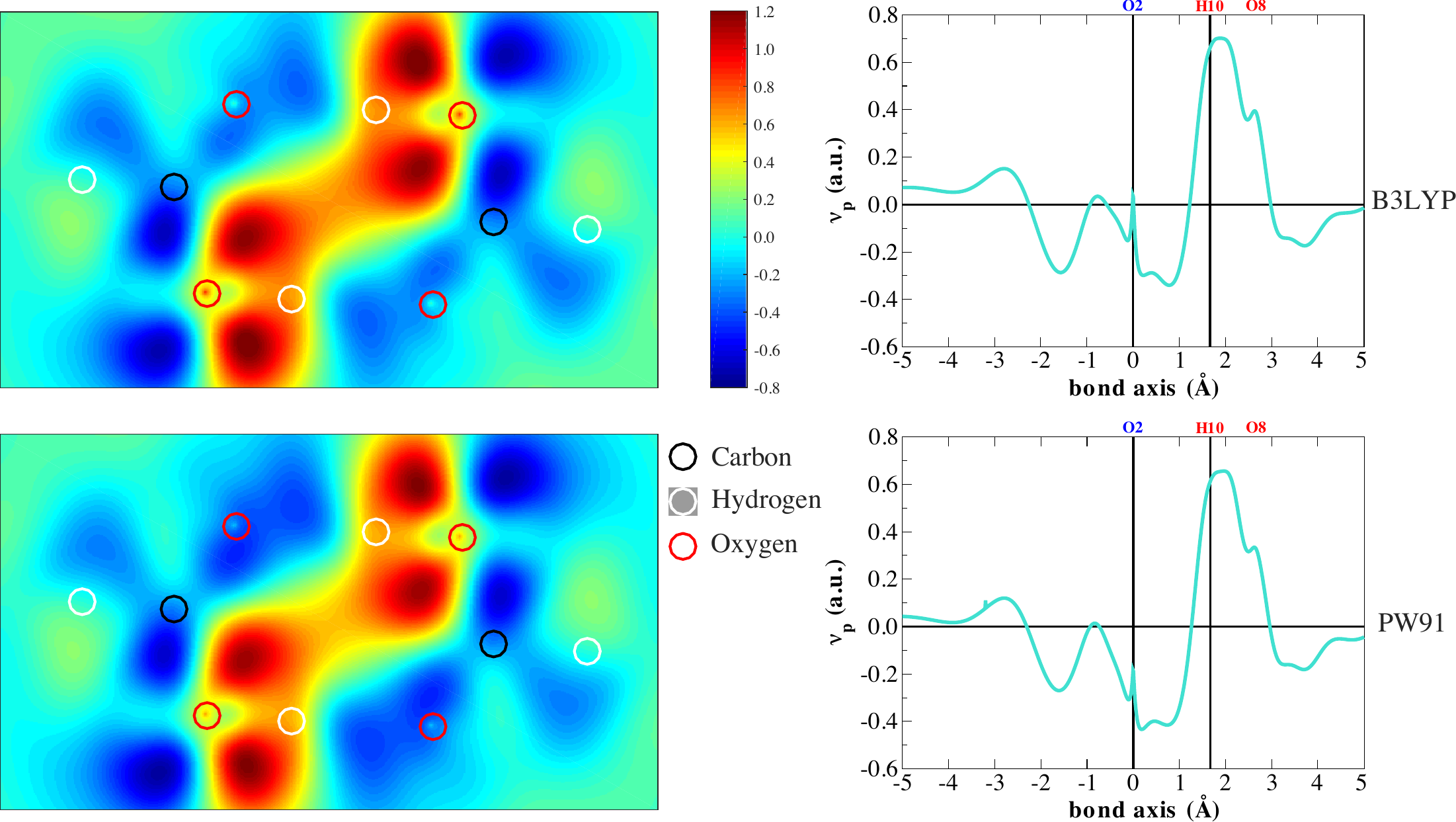} 
\caption{Partition potential, $v_p(\mathbf{r})$, for D$_1$ using B3LYP (top) and PW91 (bottom) functionals. The aug-cc-pVTZ basis set was used for all calculations. The left two plots show the $v_p(\mathbf{r})$ map on the molecular plane. The right two plots show $v_p(\mathbf{r})$ along the approximate HB line (through H atom of the donor and O atom of the acceptor).}
\label{f:rho_vp_plots}
\end{figure}

\subsection{PDFT and descriptors of weak chemical binding}

Farf\'an \textit{et al.} \cite{fad} established a hierarchy of strengths of intermolecular contacts based on a combined view that includes topological analyses of the electron densities and orbital interactions. Their results suggest that the strengths of hydrogen bonding in (FA)$_2$ decrease as C=O $\cdots$ H--O > H--O $\cdots$ H--O > C=O $\cdots$ H--C > H--O $\cdots$ H--C.  Table \ref{t:quant} lists several chemical descriptors adapted from the work of Farf\'an \textit{et al.} \textit{et al.}\cite{fad} supplemented with quantities derived from our own PDFT results. \\  

\begin{sidewaystable}
\centering
\caption{Descriptors or chemical bonding for hydrogen bonds in the formic acid dimers. WBI: Wiberg bond indices, $n$ is the electron density and $\nabla^2 n$ is its Laplacian, $\mathcal{H}$ is the total energy density, all evaluated at the bond critical points. All quantities listed in atomic units except for orbital interaction energies, $E^{(2)}_{ov}$, which are given in kcal/mol. Max ($R$) and Min ($L$) refer to the points of maximum and minimum density deformation, extracted from Figure \ref{f:dens_diff_plots}. }
\label{t:quant}
\begin{tabular}{|c|c|ccccccc|} 
\hline 
\hline
Interaction & Dimer & WBI & $n\left(\mathbf{r}_c\right)$  & $\mathcal{H}\left(\mathbf{r}_c\right)$ & $\nabla^2 n\left(\mathbf{r}_c\right)$ &   $E^{(2)}_{ov}$ & Max ($R$) & Min ($L$) \\
\hline
C=O $\cdots$ H--O  & D$_1$ & 0.0918 & 0.0398 & -0.0024 & 0.1288 &  21.36 & 0.048 & -0.170\\
C=O $\cdots$ H--O  & D$_2$ & 0.0711 & 0.0329 &  0.0001 & 0.1143 &  17.62 & 0.047 & -0.120\\
C=O $\cdots$ H--O  & D$_3$ & 0.0367 & 0.0226 &  0.0016 & 0.0816 &  6.86  & Non-Linear & Non-Linear\\
H--O $\cdots$ H--O & D$_4$ & 0.0282 & 0.0248 &  0.0021 & 0.0972  & 7.39  & 0.021  & -0.070\\
H--O $\cdots$ H--O & D$_3$ & 0.0187 & 0.0211 &  0.0021 & 0.0829 &  3.96  & 0.016 & -0.060\\
C=O $\cdots$ H--C  & D$_2$ & 0.0069 & 0.0108 &  0.0011 & 0.0370  & 1.05& Non-Linear & Non-Linear\\
C=O $\cdots$ H--C  & D$_4$ & 0.0044 & 0.0094 &  0.0008 & 0.0301  & 0.77& Non-Linear & Non-Linear\\
\hline
\end{tabular}
\end{sidewaystable}

The criteria used by Farf\'an \textit{et al.} \textit{et al.} \cite{fad}, listed in Table \ref{t:quant}, may be dissected as follows

\begin{enumerate}
 \item Wiberg bond indices are directly linked to the strength of the chemical interactions
 \item Larger electron densities at bond critical points result in stronger interactions
 \item The sign and magnitude of the total energy density at bond critical points, $\mathcal{H}\left(\mathbf{r}_c\right) = \mathcal{G}\left(\mathbf{r}_c\right) + \mathcal{V}\left(\mathbf{r}_c\right)$, are indications of the strengths and nature of the interactions because if $\mathcal{H}\left(\mathbf{r}_c\right)<0$ the potential (attractive) energy dominates leading to a concentration of electron charge in the vicinities of the BCP, strengthening the interaction, with an opposite effect for $\mathcal{H}\left(\mathbf{r}_c\right)>0$ when the kinetic (repulsive) energy dominates.
 \item The sign of the Laplacian of the electron density at bond critical points tells apart local maxima and minima. Thus, negative Laplacians (local maxima) indicate local concentration of charge around the critical points, which suggests stronger interactions. Positive Laplacians (local minima) indicate local depletion of charge, suggesting weaker interactions.
 \item Orbital interaction energies calculated via second order perturbation theory on the Fock matrix represented by the NBO basis reveal the mechanism for the formation of hydrogen bonds and are directly correlated to the strength of the interaction. In all (FA)$_2$ cases, charge transfer from a lone pair in an oxygen atom in either a carbonyl or in a hydroxyl group to a neighboring antibonding orbital is identified as the culprit. The orbital charge transfer maybe represented in a general way as $n_{\rm O} \rightarrow \sigma^{\star}_{\rm X-H}$ with X= O, C. 
\end{enumerate}

\begin{figure}
\centering
\includegraphics[scale=0.6]{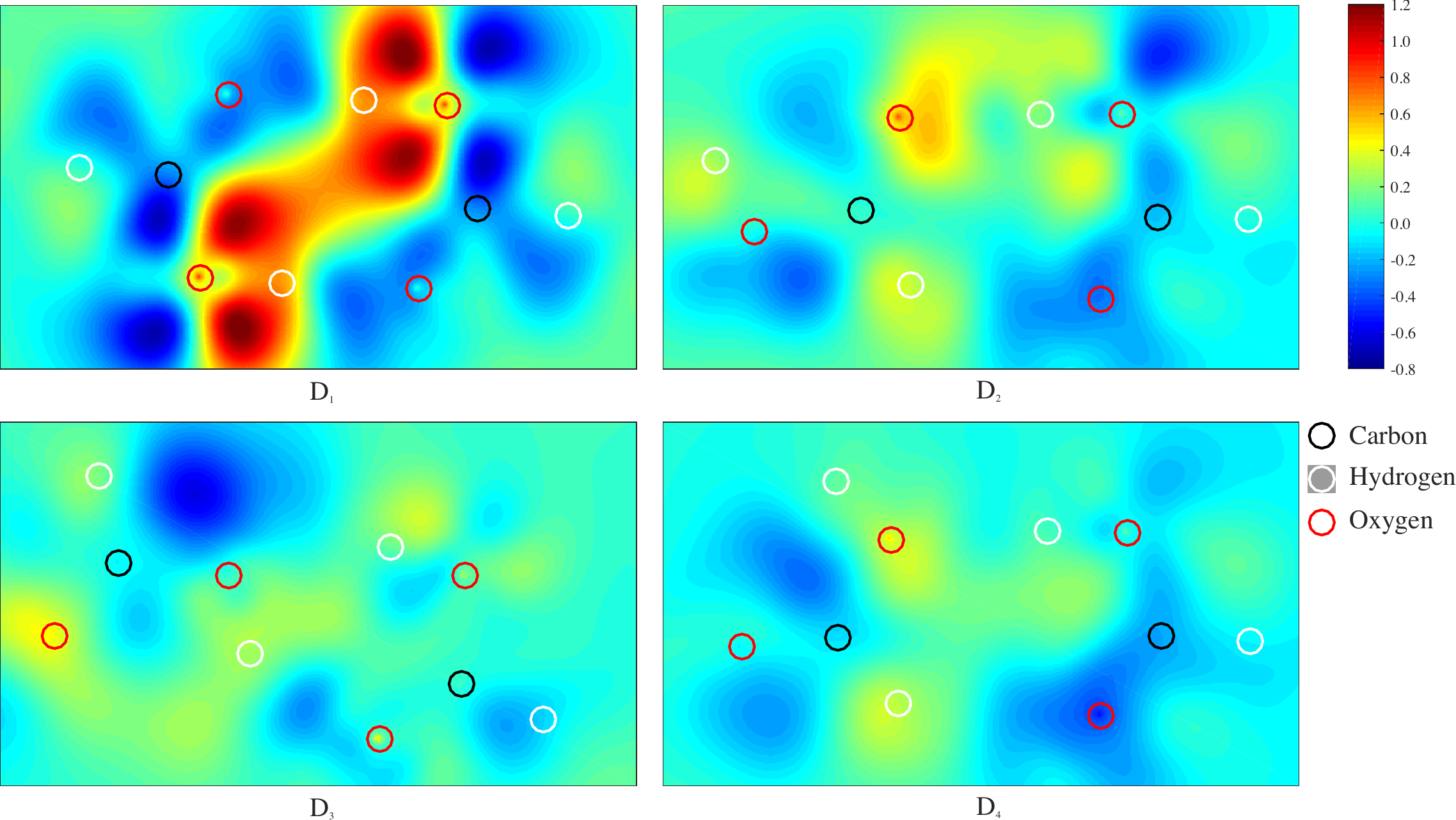} 
\caption{Partition potential, $v_p(\mathbf{r})$, for dimers for D$_1$ $\rightarrow$ D$_4$ (B3LYP/aug-cc-pVTZ).}
\label{f:vp_plots}
\end{figure}

As is clear from Table \ref{t:quant}, the low-density deformations are consistent with the hierarchy of strengths of hydrogen bonds in (FA)$_2$ established by Farf\'an \textit{et al.} \cite{fad}

\section{CONCLUDING REMARKS}

Contrary to our original expectation when we started this study, we showed that the partition potential is \textit{not} transferable between systems with similar types of hydrogen-bonding. The result highlights the nonlocal character of $v_p(\mathbf{r})$ in contrast to the local features of density deformations of the individual fragments, which \textit{are} largely transferable. In practical calculations, we should take advantage of the fact that fragment calculations can be done locally while still preserving the global features of the partition potential. We also highlight the intuitive yet nontrivial observation that large binding energies correspond to large preparation energies, and that the strength of the partition potential is correlated with the overall stability of the dimer, as made obvious by our graphical abstract. \\

\section*{acknowledgements}
We acknowledge support from the Universidad de Antioquia - Purdue grant No. PURDUE14-2-02. A.W. and Y.O. acknowledge support from the National Science Foundation CAREER program under
Grant No. CHE-1149968. 

\bibliographystyle{spphys}
\bibliography{fa2-pdft}

\end{document}